\documentstyle[12pt,psfig]{article}
\begin{document}






\newcommand{\bfrho} {\mbox{\mbox{\boldmath$\rho$}}}

\newcommand{\bftau} {\mbox{\mbox{\boldmath$\tau$}}}

\newcommand{\bfpi} {\mbox{\mbox{\boldmath$\pi$}}}
\newcommand{\bfgamma} {\mbox{\mbox{\boldmath$\gamma$}}}
\newcommand{\bfomega} {\mbox{\mbox{\boldmath$\omega$}}}

\newcommand{\fraca} {\mbox {$ \frac{1}{2}       $}}
\newcommand{\fracb} {\mbox {$ \frac{3}{2}       $}}

\newcommand{\xn} {\mbox {$ x_{N}     $}}
\newcommand{\un} {\mbox {$ u_{N}    $}}

\newcommand{\vlambda} {\mbox {$ \frac{1}{4} \lambda (\sigma^{2} -
\sigma_{0}^{2} )^{2}      $}}

\newcommand{\kf} {\mbox {$ k_{F}       $}}

\newcommand{\stressenergy} {\mbox {$ 
\calT_{\mu \nu} = -g_{\mu \nu} \calL  + \sum_{\phi} \frac{\partial \calL}
{\partial (\partial^{\mu} \phi)} \partial_{\nu} \phi $}}

\newcommand{\mstar}[1] {\mbox {$ m^{\star}_{#1}      $}}

\newcommand{\ef}[2] {\mbox {$ (\kf^{2} + \mstar{#2}^{2})^{#1 /2}  $}}

\newcommand{\ek}[2] {\mbox {$ (k^{2} + \mstar{#2}^{2} )^{#1 /2}  $}}

\newcommand{\msigma} {\mbox {$ m_{\sigma}      $}}

\newcommand{\mN} {\mbox {$ m_{N}      $}}

\newcommand{\mB} {\mbox {$ m_{B}      $}}

\newcommand{\munu} {\mbox {$ \mu \nu       $}}

\newcommand{\calT} {\mbox {$ {\cal T}       $}}

\newcommand{\calL} {\mbox {$ {\cal L}       $}}

\newcommand{\Lint}[1] {\mbox {$ \bar{\psi}_{#1} [ g_{\omega }
\gamma_{\mu} \omega^{\mu} + \frac{1}{2} g_{\rho }  \gamma_{\mu}
 \bftau\! \cdot\! \bfrho^{\mu} \cdots  ] \psi_{#1}  $}}

\newcommand{\Lsigma}[1] {\mbox {$ 
\bar{\psi }_{#1} [i \gamma_{\mu} \partial^{\mu} -g( \sigma
+i \gamma_{5}\bftau\! \cdot\! \bfpi) ] \psi_{#1}  $} }

\newcommand{\Lmeson} {\mbox {$
\frac{1}{2} (\partial_{\mu} \sigma \partial^{\mu} \sigma
  + \partial_{\mu}\bfpi\! \cdot\! \partial^{\mu}\bfpi)
  - \frac{1}{4} \lambda (\sigma^{2}+\bfpi\! \cdot \!
  \bfpi- \sigma_{0}^{2})^2 $} }


\newcommand{\EM}{\sqrt{k^2+m^{2}_{M}}}
\newcommand{\LLL}[1]{ {\cal L}_{#1}^{0} }
\newcommand{\U}{\mbox{$
 \frac{1}{3} b m_{n} (g_{\sigma} \sigma_{0})^{3}
+ \frac{1}{4} c(g_{\sigma}\sigma_{0})^{4}  $}}
\newcommand{\Upm}{\mbox{$ -\;b m_n (g_{\sigma} \sigma_{0})^2
-c (g_{\sigma} \sigma_{0})^3  $}}
\newcommand{\M}{\mbox{$ \sqrt{ k^{2}+(m_{B}-g_{\sigma B}\sigma_{0})^{2}
} $}}
\newcommand{\m}{\mbox{$ \sqrt{ k^{2}+m_{\lambda}^{2} }  $}}
\newcommand{\ke}[1]{\mbox{$ \frac{1}{2} m_{#1}^{2} #1^{2}  $}}
\newcommand{\keo}{\mbox{$\frac{1}{2} m_{\omega}^2 \omega_{0}^{2} $}}
\newcommand{\kerho}{\mbox{$\frac{1}{2} m_{\rho}^2 \rho_{03}^{2} $}}
\newcommand{\Um}{\mbox{$
-\; \frac{1}{3} b m_{n} (g_{\sigma} \sigma_{0})^{3}
- \frac{1}{4} c(g_{\sigma}\sigma_{0})^{4}  $}}

\newcommand{\meff}{m_{B}-g_{\sigma B}\sigma}
\newcommand{\coup}[1]{\mbox{$ (g_{#1}/m_{#1})^{2} $}}

\newcommand{\lsigma}{\mbox{$\frac{1}{2}(\partial_{\mu} \sigma
\partial^{\mu} \sigma - m_{\sigma}^{2} \sigma^{2})$  }}

\newcommand{\lomega}{\mbox{$ - \: \frac{1}{4} \omega_{\mu \nu} 
\omega^{\mu \nu} +\frac{1}{2} m_{\omega}^{2} \omega_{\mu} \omega^{\mu} $}}

\newcommand{\lrho}{\mbox{$ 
- \: \frac{1}{4}\bfrho_{\mu \nu}\! \cdot\! \bfrho^{\mu \nu} 
+ \frac{1}{2} m_{\rho}^{2}\bfrho_{\mu}\! \cdot\! \bfrho^{\mu} $}}

\newcommand{\omunu}{\mbox{$ \omega_{\mu \nu} = 
\partial_{\mu} \omega_{\nu}
- \partial_{\nu} \omega_{\mu} $}}

\newcommand{\nbaryon}[1] { \mbox{$ 
    \bigl(  \exp [(\epsilon_{B}(k) {#1})/T] +1  \bigr) ^{-1} $ } }

\newcommand{\nmeson}[2] {\mbox{$
    \bigl(  \exp [(\omega_{{#2}} (k) {#1})/T] -1 \bigr) ^{-1} $ } }

\newcommand{\D} {\mbox{$ {\cal D}(k,\omega) $} }

\newcommand{\Dm} {\mbox{$ {\cal D}^{-1}(k,\omega) $} }

\hyphenation{
temp-erature temper-ature cor-respond corres-pond
}


\newcounter{sctn}
\newcounter{subsctn}[sctn]
\newcommand{\sctn}[1]{~\\ \refstepcounter{sctn} {\bf \thesctn~~ #1} \\ }
\newcommand{\subsctn}[1]
{~\\ \refstepcounter{subsctn} {\bf \thesctn.\thesubsctn~~ #1}\\}


\newcommand{\msun}{M_{\odot}}
\newcommand{\eos}{equation of state~}
\newcommand{\eoss}{equations of state~}
\newcommand{\eossp}{equations of state}
\newcommand{\eosp}{equation of state}
\newcommand{\Eos}{Equation of state~}
\newcommand{\Eoss}{Equations of state~}
\newcommand{\beqn}{\begin{eqnarray}}
\newcommand{\eeqn}{\end{eqnarray}}
\newcommand{\nonum}{\nonumber \\}
\newcommand{\walecka}{$\sigma-\omega$~}


\newcommand{\dateofdoc}{July 28, 1995}

\newcommand{\lbl}{\begin{flushright} LBL-37566\\[7ex] \end{flushright}}

\newcommand{\tit}
{Crystalline Structure in the Confined-Deconfined Mixed Phase: Neutron
Stars}

\newcommand{\doe}
{This work was supported by the 
Director, Office of Energy Research, 
Office of High Energy
and Nuclear Physics,
Division of Nuclear Physics,
of the U.S. Department of Energy under Contract
DE-AC03-76SF00098.}

\newcommand{\kade} {Supported by a fellowship from the Max Kade
Foundation.}
\newcommand{\autha} {N. K. Glendenning}
\newcommand{\authb} {S. Pei \\[2ex]}

\newcommand{\adra}
{{\em Nuclear Science Division and\\
Institute for Nuclear \& Particle Astrophysics\\
Lawrence Berkeley National Laboratory\\
University of California\\
One Cyclotron Road\\
Berkeley, California 94720}\\[2ex]  }

\newcommand{\adrb}
{{\em Beijing Normal University\\
Department of Physics\\
Beijing 199875\\
P. R. China}\\[2ex]}


\begin{titlepage}
\lbl

\begin{center}
\begin{Large}

\renewcommand{\thefootnote}{\fnsymbol{footnote}}
\setcounter{footnote}{1}

\tit {\footnote{\doe}}\\[7ex]

\end{Large}

\begin{large}
\renewcommand{\thefootnote}{\fnsymbol{footnote}}
\setcounter{footnote}{2}
\autha \\[2ex]
\end{large}
\adra
\begin{large}
\authb
\end{large}
\adrb
\dateofdoc \\[10ex]
\end{center}

\begin{quote}
\begin{center}
{\bf 
Invited Contribution to\\
Heavy Ion Physics, {\bf 1} (1995) 323\\
Acta Physica Hungarica New Series\\
Devoted to the Memory of\\[1ex]
\Large{Eugene Wigner}.}
\end{center}
\end{quote}
~~\\[3ex]

\begin{center}
\end{center}
\end{titlepage}

\renewcommand{\thefootnote}{\arabic{footnote}}
\setcounter{footnote}{0}


\begin{center}

\begin{Large}
\tit
\end{Large}
\begin{large}
\\[6ex]
\autha~~~ and~~~ \authb
~\\[3ex]
\end{large}
\end{center}
\begin{quote}
We review the differences in first order phase transition of
single and multi-component systems, and then discuss the crystalline
structure expected to exist in the mixed confined deconfined 
phase of hadronic matter. The particular context of neutron stars
is chosen for illustration. The qualitative results are general
and apply for example to the vapor-liquid transition in subsaturated
asymmetric nuclear matter.
\end{quote}

\normalsize


\section{Introduction}

First order phase transitions are very familiar only in one-component
substances such as water. As is well known, on an isotherm the pressure
remains constant as do all internal properties such as density  and chemical 
potential,  for all proportions of the two phases, gas and liquid,
in equilibrium. These  characteristic  properties of {\sl all}
single-component 
substances are unique to them, and are not
at all general. What is not familiar is that
the precise converse of the above
properties holds when the
substance has more than one independent component. This has unique consequences in certain situations,
such as in the presence of a gravitational field. 
More than that, when one of the independent components is electrically
charged, the two phases in equilibrium may form a Coulomb lattice
of the rare phase immersed in the dominant one. We have proven these
properties in great detail  and generality elsewhere \cite{glen91:a}.
Our aim here is to briefly recapitulate the physical reason
for the different behavior of  a first order
phase transition in single- and multi-component substances,
and then to compute the varying geometry of the crystalline
structure as a function of proportion of the phases in equilibrium.
We shall do this in the context of the confined-deconfined
phase transition in neutron star matter, --  matter that
is charge neutral and in equilibrium with respect to all baryon
and quark species. The results would be qualitatively similar for the
liquid-vapor transition in sub-saturated nuclear matter.

\section{Degrees of Freedom in  Multi-component
System}

We stated above that in the mixed phase of a multi-component substance
all internal properties of each phase and their common pressure vary
as the proportion of the phases.
Let us see why this is so, first by considering the physics rather than
the mathematics.
 Consider a substance composed of
 two  conserved `charges' or independent
 components, -- Q of one kind, B of the other.
 In the case of a neutron star, these could denote the net electric charge
 number (in units of e) and baryon charge number.
 Let the substance be closed and in a heat bath. 
  Define their concentration,
 \beqn  
 r=Q/B \,.
 \eeqn  
Is this ratio fixed? One would certainly think so since $Q$ and $B$
are fixed.
But the ratio
is fixed {\sl only} as long as the system remains in one pure phase or the
other! When in the mixed phase the concentration in each of the
regions of one phase or the other may be different and they are restricted
{\sl only} by the conservation on the total numbers,
\beqn r_1 =Q_1/B_1,~~~
r_2=Q_2/B_2,~~~~~~
(Q_1+Q_2=Q,~~~ B_1+B_2=B) \,.
\eeqn
{\sl If the internal forces can lower the energy} of the system
 by rearranging the concentration, they will do so.
 The essential point is that conservation laws in chemical 
 thermodynamics are global, not local.

The above observations allow us to prove easily that all properties of 
each phase in equilibrium with the other will vary according to the
proportion of the phases.
Consider the system at the density or pressure
where the neutron star matter has just begun to condense some quark matter.
There is little scope for the internal forces to optimize the
concentrations, $r_1,r_2$, in the two phases,
since the small quantity of quark matter
can neither receive nor donate much of either charge. However,
at higher density or pressure, the proportion of the two phases will become
more comparable, and the internal forces now have more scope to
optimize the concentrations in the two phases, always consistent with
overall conservation of the two charges. From this observation, we learn:
  {\sl For a first order phase
  transition in a multi-component system, the nature 
  of each phase in equilibrium changes with the
proportion of the phases and since the total energy is now the volume
proportion of the energy density of the two phases, each of which varies
with the proportion, the derivative with respect to volume is no longer
a constant. Therefore
the pressure also varies as the proportion of phases!}

The mathematical proof of the above properties is not nearly so illuminating
as the physical verbal proof above, but we give it for completeness.

The Gibbs condition for phase equilibrium is that the chemical potentials
$\mu_b,~\mu_q$ corresponding to 
$B$ and $Q$, temperature T and the pressures in the two phases
be equal,
\beqn
p_1(\mu_b,\mu_q,T)=p_2(\mu_b,\mu_q,T)
\label{pres}
\eeqn
As discussed,  the condition of {\sl local}  conservation
is stronger than required.
We apply the weaker condition of {\sl global}
  conservation,
\beqn
& &<\rho>\equiv (1-\chi) \rho_1(\mu_b,\mu_q,T)+ \chi \rho_2(\mu_b,\mu_q,T)=
 B/V\, , \label{dens} \\
& & (1-\chi) q_1(\mu_b,\mu_q,T)+ \chi q_2(\mu_b,\mu_q,T)=Q/V\, ,
~~~~\chi=V_2/V\,.
\label {neut}
\eeqn
Given a temperature,
  the above three equations serve to 
determine the two independent chemical potentials and $V$
 for a {\sl specified}
volume fraction $\chi$ of phase `2' in equilibrium with phase `1'.
We note that the condition of global conservation expressed by
(\ref{dens}) and 
(\ref{neut})
is compatible, together with (\ref{pres}), 
with the number of unknowns
to be determined. It would {\sl not} be possible to satisfy Gibbs conditions
if {\sl local} conservation were demanded, for that would replace (\ref{neut})
by  {\sl two} equations, such as
$q_1(\mu_b,\mu_q,T)=Q_1/V_1,~~q_2(\mu_b,\mu_q,T)=
Q_2/V_2$,
and the problem would be over determined. 

In systems possessing only one conserved charge, 
the pressure equation defines uniquely the corresponding  chemical potential
for phase equilibrium.
 In that case the energy densities of each phase
are also determined as  unique
values and like the pressure are {\sl independent} of the
proportion of the phases in equilibrium. In contrast with this,
for two or more
conserved charges and corresponding chemical potentials, the situation
is quite different. Through (\ref{pres},\ref{dens},\ref{neut}) 
the chemical potentials
obviously depend on the proportion, $\chi$,
of the phases in equilibrium, and hence so
also all properties that depend on them, the  energy densities, baryon and
charge densities
of each phase. and the {\sl common} pressure.
This remarkable and little known property of first order
phase transitions with more than one conserved charge
and the role played by the microphysics or internal forces
is discussed in detail elsewhere
\cite{glen91:a,glent94:e}.
It will be observed that the above discussion is completely general,
and must apply to many systems, in particular, to the confined-deconfined
phase transition at high density and equally to the so-called
liquid-vapor transition in nuclear matter at sub-saturation density.
For both systems the symmetry energy  is the driving force, and 
clearly the results here for two component systems hold when $Z \neq N$.
For the {\sl special} case of equality however, the driving force is
absent, -- both phases are already symmetric, -- and the pressure
would be constant throughout the mixed phase. But only when $N=Z$!

\section{Internal Forces}

By the above discussion we understand that the internal force(s)
can exploit the degree(s) of freedom available in rearranging
concentrations of conserved quantities while conserving them globally
and lowering the energy. Let us look now at a specific example,
neutron star matter which is charge neutral and in chemical equilibrium.
Stars must be neutral because they are bound by gravity and net charge
would reduce their binding, it being also long-ranged.
Since pure neutron matter is beta unstable, neutron star matter
will be composed of various particles of different
charges, --  neutrons, protons, leptons,
perhaps
hyperons and quarks.  The star is born with a definite number of baryons,
and soon becomes neutral. There are two conserved charges, therefore, --
electric charge and baryon number, -- and two corresponding independent
chemical potentials.

The internal force that can exploit the degree of freedom made available
by allowing neutrality to be achieved 
{\sl globally} 
and which is  closed to one   in which  {\sl local} neutrality
is artificially enforced,
is the isospin restoring
force experienced by the confined phase of hadronic matter.
It is embodied in the isospin symmetry energy in the empirical
mass formula of nuclei and nuclear matter.
The hadronic regions of the mixed phase  can arrange to be
more isospin symmetric 
  (closer equality in proton and neutron number) than in the pure phase
  by transferring
charge to the
 quark phase in equilibrium with it.
 Symmetry energy will be lowered thereby at only a small cost in rearranging
 the quark Fermi surfaces.
 Electrons play
only a minor role when neutrality can be realized among baryon charge
carrying particles. Thus the mixed phase region of the star
will have {\sl positively} charged regions of nuclear matter and 
{\sl negatively}
charged regions of quark matter.

 \section{Structure in the Mixed Phase\label{structure}}

The Coulomb interaction
will tend to break the regions of like charge into smaller ones, while
this is opposed by the surface interface energy.
Their competition will be resolved by {\sl  forming
a lattice of the rare phase immersed in the dominant
one whose  form, size and spacing 
will minimize the 
sum of surface and Coulomb
energies}. In other words, a crystalline lattice 
will be formed. Since all internal properties
of the two phases in equilibrium with each other vary with their 
proportion, so will the geometrical structure.
When quark matter is the rare phase immersed in 
confined hadronic matter, it will form droplets. At higher proportion
of quark matter, the droplets will merge to from strings and then sheets,
and then the role  in the geometric structure
of confined and deconfined phases will interchange \cite{glen91:a}.

We consider  a Wigner-Seitz cell of radius $R$ containing the rare phase object
of radius $r$
and an amount of the dominant phase that makes the cell charge neutral.
The whole medium can be considered as made of such non-interacting cells,
under the usual approximation of neglecting the interstitial material.
As we shall see, the size of these cells is is of the order of tens
of fermis or less.  The variation of the metric over such small regions
is completely negligible (see ref. \cite{glen85:b} for the radial
behavior of the metric in typical neutron star
models), so they are locally inertial
regions and our discussion of them as if gravity is absent is justified
by the equivalence principle. 
The solution to problems involving a competition between Coulomb and surface
interface energies is universal.
We may adapt  the results of
\cite{ravenhall83:a} to write for the radius
 of the rare phase immersed in the other
and the minimum of the sum
of Coulomb and surface energies, in the case of  three geometries,
slabs, rods and drops,
\beqn
\frac{1}{r^3} &=& \frac{4\pi [q_H(\chi)-q_Q(\chi)]^2 e^2
f_d(x)}{\sigma d}\,,~~~~~d=1,2,3\,,\\
\frac{E_C+E_S}{V} &=& 6\pi x
\Bigl( \frac{[\sigma d \bigl(q_H(\chi)-q_Q(\chi)
\bigr)e]^2
 f_d(x)}
{16\pi^2}\Bigr)^{1/3} \,,
\eeqn
where, $q_H,~q_Q$ are the charge densities of hadronic and quark matter
(in units of $e$)
at whatever proportion $\chi$ being considered.
We have denoted the
volume fraction of quark matter $V_Q/V$ by $\chi$. 
The ratio of droplet (rod, slab)    to cell volume is called,
\beqn
x= (r/R)^d\,.
\eeqn
It is related to $\chi$ by,
\beqn
\chi = (r/R)^d\equiv x~~~~({\rm hadronic~matter~background})\,,
\eeqn
when hadronic matter is the background (ie. dominant)
phase. The quark droplets (rods, slabs)
have radius $r$ and the spacing between centers is $R$,
with $d=1,2,3$ corresponding 
to slabs, rods and drops, respectively. In the case
of drops or
rods, $r$ is their radius and $R$ the half distance between centers while for
slabs, $r$ is the half thickness.
In the opposite
situation where quark matter is the background,
\beqn
1-\chi = (r/R)^d\equiv x~~~~({\rm quark~matter~background})\,,
\eeqn
is the fraction of hadronic matter which assumes the above geometric
forms.

The function $f_d(x)$ is given in all three cases by,
\beqn
f_d(x) =\frac{1}{d+2}
\Bigl[\frac{1}{(d-2)}(2-d x^{1-2/d}) + x \Bigr]\,,
\eeqn
where the
apparent singularity for $d=2$ is well behaved and has the
correct value,
\beqn
 \stackrel{ {\textstyle {\rm limit} } } { _{d\rightarrow 2} }
 \frac{1}{(d-2)}(2-d x^{1-2/d})\longrightarrow -[1+\ln x ]\,.
 \eeqn
We have supposed that the electrons are uniformly distributed
throughout the
mixed phase
whether quark or hadronic regions, and hence they do not appear in the
above.
In fact, we shall find that electrons are almost totally absent 
from the mixed phase.

 What can we say of the surface tension? This is a very difficult
 problem to solve. Obviously it should be self-consistent with the two
 models of matter, quark and hadronic, in equilibrium with each other.
 This latter feature arises because of the fact that, unlike simple
 substances like water and vapor, the densities of each phase change as their
 proportion does \cite{glen91:a,glen91:d}.
 So the surface energy is not a constant.
 Following our deduction that a Coulomb lattice should exist in the
 mixed hadron-quark phase \cite{glen91:a,glen91:d}, Heiselberg,
 Pethick and Staubo have investigated the dependence
 of the geometrical structure
 on the surface tension 
  \cite{heiselberg93:a}.
 They adopted a selection of values from various sources, none of them computed
  self-consistently, for this is an extremely hard problem. 

 Gibbs studied the problem of surface energies,
 and as a gross approximation, one can deduce that it is
 given by the difference in energy densities
 of the substances in contact times a length scale typical
 of the surface thickness \cite{myers85:a}, in this case of the
 order of the strong interaction range, $L= 1$ fm.
 In other words, the surface interface energy should depend on the
 proportion of phases in phase equilibrium, just as everything else does.
 \beqn
 \sigma =  {\rm const} \times[\epsilon_Q(\chi) -\epsilon_H(\chi)]\times L\,,
 \label{avenergy}
 \eeqn
 where $\chi$ is the volume proportion of quark phase.
 The constant should be chosen so that the structured phase lies below the
 unstructured one. Heiselberg et al found this energy difference to be 
 about 10 MeV. We choose the constant accordingly. 
 
 It will be understood
 from the formulae written above that the structure size, whether
 drops, rods or slabs, and the sum of surface and Coulomb energies
 scale with the surface energy coefficient as
 $\sigma^{1/3}$ independent of geometry. 
 Therefore the location in the star where the geometry changes from one form
 to another is independent of $\sigma$.

\section{Bulk Description of the Phases\label{bulk}}

The geometrical structure of the mixed phase occurs against the background
of the bulk structure, at least to good approximation. The energy
and pressure are of course dominated by the bulk properties of matter.
We outline briefly how to handle this part of the problem.
It has been discussed in detail elsewhere.

For the confined hadronic phase we use 
the covariant
Lagrangian,
\begin{eqnarray}
{\cal L} & = &  
\sum_{B} \overline{\psi}_{B} (i\gamma_{\mu} \partial^{\mu} - m_{B}
+g_{\sigma B} \sigma  - g_{\omega B} \gamma_{\mu} \omega^{\mu}
- \fraca g_{\rho B} \gamma_{\mu} \bftau \cdot \bfrho^{\mu} )
{\psi}_{B} \nonumber  \\
&   &+\:  \lsigma
  \lomega  \nonumber\\[2ex]
&   &  \lrho 
  \:\Um  \nonumber\\[2ex]
&   &  + \sum_{e^{-},\mu^{-}} 
\overline{\psi}_{\lambda} \bigl(i\gamma_{\mu}
  \partial^{\mu} - m_{\lambda} \bigr) \psi_{\lambda}\,.
  \label{lagrangian}
\end{eqnarray}
We regard it as an effective theory to be solved at the 
mean field level, and with coupling constants adjusted,
as described below, to nuclear matter properties.
The baryons, $B$ are coupled to the $\sigma, \omega, \bfrho$
mesons.
The sum on $B$ is over all the charge states of the lowest baryon
octet, ($p,n,\Lambda,\Sigma^{+},\Sigma^{-}, \Sigma^{0},
\Xi^{-}, \Xi^{0}$) as well as the $\Delta$ quartet. However
the latter are not populated up to the highest density in neutron
stars, nor are any other baryon states save those of the lowest
octet for reasons given elsewhere \cite{glen85:b}. 
The last term represents the free lepton Lagrangians.
How the theory can be solved in the mean field 
approximation for the ground state of charge neutral
matter in general beta equilibrium (neutron star matter)
is described fully in
ref. \cite{glen85:b}.

There are five  constants here that are determined
by the properties of nuclear matter,
three that determine the nucleon couplings to the
scalar, vector and vector, iso-vector mesons, $g_\sigma/m_\sigma,
g_\omega/m_\omega, g_\rho/m_\rho$,
and two that determine the scalar self-interactions, b,c.
The nuclear properties that define their values
are the saturation values of the binding energy, baryon density,
symmetry energy coefficient, compression modulus and
nucleon effective mass. 
The hyperon couplings are not relevant to the ground state
properties of nuclear matter but information about them
can be gathered from levels in hypernuclei, the binding of the $\Lambda$
in nuclear matter, and from neutron star masses \cite{glen91:c}. 
We shall assume that
all hyperons in the octet have the same coupling
as the $\Lambda$.
They are expressed as a ratio to the above mentioned
nucleon couplings,
\beqn
x_\sigma=g_{H\sigma}/g_{\sigma},~~~~
x_\omega=g_{H\omega}/g_{\omega},~~~~
x_\rho=g_{H\rho}/g_{\rho}.
\label{ratio}
\eeqn
The first two are related to the $\Lambda$ binding by a relation derived
in \cite{glen91:c} and the third can be taken equal to the second
by invoking vector dominance.
We adopt the value of $x_\sigma=0.6$ and corresponding $x_\omega$ 
taken from 
 \cite{glen91:c}.

The chemical potentials of all hadrons are given by,
\beqn
\mu_B  =b_B \mu_n-e_B\mu_e \,,
\label{equil}
\eeqn
where $b_B$ and $e_B$ are the baryon and electric charge numbers of the
baryon state B, and $\mu_n$ and $\mu_e$ are independent
chemical potentials for unit baryon number and unit negative electric charge
number
(neutron and electron respectively).

The  values of nuclear matter properties
are the binding, $B/A=-16.3$ MeV, saturation density, $\rho_0=0.153$ fm$^{-3}$,
and symmetry energy coefficient, $a_{{\rm sym}}=32.5$ MeV, 
$K=240$ MeV, $\mstar{{\rm sat}}/m=0.78$.

To describe quark matter
we use a simple version of the bag model for finite quark masses
and
 $T=0$ \cite{glen91:a}.
Because of the long time-scale, strangeness is not conserved in
a star. The quark chemical potentials  for a system in chemical
equilibrium are therefore related to those
for baryon number and electron by
\beqn
\mu_u=\mu_c=
    \frac{1}{3} (\mu_n-2 \mu_e),~~~~~~~\mu_d=\mu_s=
    \frac{1}{3} (\mu_n+\mu_e).
    \eeqn

Solving the models of confined and deconfined phases, in both pure phases and
in the mixed phase, we can compute the composition of charge-neutral,
beta-stable neutron star matter. It is shown in Fig.\ \ref{pops240}.
Note the saturation of the leptons as soon as quark matter appears.
At this stage, charge neutrality is achieved more economically on baryon charge
carrying particles, since the star has a definite baryon number.
We note the transition from pure hadronic to mixed phase occurs at
the rather low density of about $2\rho_0$, as was found also by several
other authors \cite{heiselberg93:a,pandharipande94:a}.

\begin{figure}[tbh]
\vspace{-.5in}
\begin{center}
\leavevmode
\centerline{ \hbox{
\psfig{figure=ps.wigner1,width=2.5in,height=3in}
\hspace{.6in}
\psfig{figure=ps.wigner2,width=2.5in,height=3in}
}}
\begin{flushright}
\parbox[t]{2.7in} { \caption{ \label{pops240} Baryon, lepton 
and quark populations
in charge neutral, beta-stable neutron star matter,
as a function baryon density $<\rho>$.
In the mixed phase region, the quark densities refer to the their values
averaged over the volume of a Wigner-Seitz cell
 and similarly for the baryons.
}} \ \hspace{.4in} \
\parbox[t]{2.7in} { \caption { \label{cry} The bulk energy density of the 
hadronic and quark
phases in equilibrium as a function of local volume proportion of the
quark phase, $\chi=V_Q/V$, the surface energy coefficient,
$\sigma(\chi)$, proportional
to the difference of the above, and the sum of Coulomb and surface energies.
}}
\end{flushright}
\end{center}
\end{figure}

\section{Varying Crystalline Structure}

We are now in a position to compute the geometrical structures,
their sizes and spacings as they vary from one radial point to another
throughout the mixed phase region.
Our purpose is to demonstrate the extreme dependence
of the structure 
of the crystalline region as a function of 
proportion of phases or equivalently density or  pressure.

In Fig.\ \ref{cry} we show some of the ingredients from the bulk
calculation that enter the computation of the structure as laid down
in section \ref{structure}. 
It is noteworthy how the energy density,
of each phase varies
 throughout
the mixed phase region
as a function of the volume fraction of quark matter, just as
we showed above must be the case in general.
Therefore the total energy density,
\beqn
\epsilon(\chi)= (1-\chi)\epsilon_H(\chi)+ \chi \epsilon_Q(\chi)\,,
\eeqn
is a {\sl non-linear} function of proportion (or volume).
As a consequence, the
pressure varies throughout the mixed phase.
This is in contrast to a simple substance, one with only one conserved charge,
in which the density of each phase in equilibrium  remains constant
as well as the pressure.
It is also worth noting that the bulk energy densities of the
confined and deconfined phase are about two orders of magnitude greater than
the sum of the energy densities of the Coulomb and surface interface energy.
This justifies the two part approach to the problem, of computing the
bulk properties and then against this background, the geometrical structure
imposed by the surface and Coulomb energies. As already noted, the total
charge 
in a Wigner-Seitz cell is zero, so  the Coulomb force is shielded by the 
lattice arrangement of the rare phase immersed in the dominant.
To illustrate the rearrangement of the electric charge concentration
between the quark and baryonic regions of the mixed phase, we show the
charge density in each region, and the electron charge density, assumed to
be uniform throughout the Wigner-Seitz cell, as functions of the
proportion of quark matter in Fig.\ \ref{chiq}. It is interesting to see that
quark matter, which in the absence of baryonic matter ($\chi=1$) is 
charge neutral, carries a high negative charge density when there is little
of it and it is in equilibrium with baryonic matter. The latter
acquires an ever increasing density as the quantity of quark matter,
with which it can balance electric charge, grows. This illustrates how 
effectively the symmetry driving force acts to optimally rearrange
charge.
\begin{figure}[tbh]
\vspace{-.5in}
\begin{center}
\leavevmode
\centerline{ \hbox{
\psfig{figure=ps.wigner3,width=2.5in,height=3in}
\hspace{.6in}
\psfig{figure=ps.wigner4,width=2.5in,height=3in}
}}
\begin{flushright}
\parbox[t]{2.7in} { \caption {\label{chiq} The charge densities in the
mixed phase carried by regions of quark and hadronic matter, as well as leptons
which permeate all regions in our approximation. Multiplied by the
respective volumes occupied, the total charge
adds to zero.
}} \ \hspace{.4in} \
\parbox[t]{2.7in} { \caption { \label{chi1} Diameter (lower curves) and Spacing
(upper curves) of rare phase immersed in the dominant as a function
of the proportion of quark phase. Geometries are identified as drops, rods,
slabs, and composition as q (quark) or h (hadronic). Dots are
a continuous dimensionality interpolation of the discrete shapes.
}}
\end{flushright}
\end{center}
\end{figure}

As shown above, because one of the conserved quantities is the  electric
charge, having long range, an  order will be established
in the mixed phase, the size of the objects of the rare phase
and their spacing in the dominant one, being determined by the
condition for a minimum sum of Coulomb and surface energy. 
In Fig.\ \ref{chi1} the diameter D and spacing S is shown by the lower
and upper curves as a function of proportion of quark phase.
The discrete geometries are labeled and their content as quark or hadronic
by `q' or `h'. The dotted line shows a continuous dimensionality 
interpolation. It is noteworthy that at the limit of the pure phases 
corresponding to $\chi = 0 {~\rm or~} 1$, the spheres of rare
phase are of finite diameter, but spaced far apart.
The size of the objects is between 7 and 15 fm. As noted previously 
the location in $\chi$ of the geometries is independent of 
$\sigma$, but the size and spacings scale as $\sigma^{1/3}$.

We have exhibited the crystalline structure of the mixed phase of
confined and deconfined neutron star matter. What is of crucial
importance is that the mixed phase, if it had a constant pressure
for all proportions, would be absent from the star, or any gravitational field.
This is because a constant pressure region cannot support any overlaying
material, and the pressure is monotonic in a star as it is in our
atmosphere.
This squeezing out of the mixed phase
was an inadvertent feature of idealizations of all treatments
of the phase transition in neutron stars until our work. The idealizations 
were either an assumption of {\sl purely} neutron star
\cite{quark00}, or an
assumption of {\sl local} charge neutrality
\cite{quark2}. Neither is a  valid
constraint.

It is almost certain that a solid region in a pulsar will play a role
in the period glitch phenomenon, which is highly individualistic
from one pulsar to another. We have  suggested that this
high degree of individual behavior may be due to the extreme
sensitivity on stellar mass of the radial extent of the solid region
and the particular geometrical forms and sizes of the objects at the
lattice sites
\cite{glen95:c}. The sensitivity arises because of the rather flat
radial profile of the pressure and energy density in neutron stars,
so that a small change in central density and therefore a small
change in stellar mass, moves a transition pressure a considerable distance
in the radial direction in the star.

As remarked earlier, we have illustrated very general phenomena 
associated with first order phase transitions in multi-component
systems. Whether geometric structures can develop on the time scale of 
collisions between nuclei is problematic, but the non-constant
pressure in the mixed phase is likely to have consequences that
may be observable. In particular the so called plateau behavior
ascribed to phase transitions in nuclear collisions can be
present only for N=Z symmetric systems, since otherwise the
symmetry energy will have scope to act in the mixed phase.

{\bf Acknowledgements}:
       \doe.



\clearpage

\end{document}